\documentclass[conference]{IEEEtran}
\usepackage{cite}
\usepackage{subcaption}
\usepackage{xspace}
\usepackage{amsmath,amssymb,amsfonts}
\usepackage{algorithmic}
\usepackage{graphicx}
\usepackage{textcomp}
\usepackage{xcolor}
\usepackage{hyperref}
\usepackage{ifthen}
\usepackage{wrapfig}
\def\BibTeX{{\rm B\kern-.05em{\sc i\kern-.025em b}\kern-.08em
    T\kern-.1667em\lower.7ex\hbox{E}\kern-.125emX}}
\newcommand{\vts}{{\sc V2S}\xspace}
\newcommand{\Faster}{{\sc Faster R-CNN}\xspace}

\newcommand{\AlexNet}{{\sc AlexNet}\xspace}
\newcommand{\Opacity}{{\sc Opacity CNN}\xspace}

\begin{document}

\title{V2S: A Tool for Translating Video Recordings of Mobile App Usages into 
Replayable Scenarios}

\author{

\IEEEauthorblockN{Madeleine Havranek\IEEEauthorrefmark{1}, Carlos Bernal-Cárdenas\IEEEauthorrefmark{1}, Nathan Cooper\IEEEauthorrefmark{1},\\
Oscar Chaparro\IEEEauthorrefmark{1}, Denys Poshyvnayk\IEEEauthorrefmark{1}, Kevin Moran\IEEEauthorrefmark{2}}
\IEEEauthorblockA{\IEEEauthorrefmark{1}\textit{College of William \& Mary} (Williamsburg, VA, USA), \IEEEauthorrefmark{2}\textit{George Mason University} (Fairfax, VA, USA)\\mrhavranek@email.wm.edu,
cebernal@cs.wm.edu, nacooper01@email.wm.edu, \\oscarch@wm.edu, denys@cs.wm.edu, kpmoran@gmu.edu}
\and


}

\maketitle



\newcommand{\tnew}[1]{{\bf { #1 }} }


\newcommand{\ndef}{\stackrel{\rm def}{=}}




\pagenumbering{arabic}



\newboolean{showcomments}

\setboolean{showcomments}{true}

\ifthenelse{\boolean{showcomments}}
  {\newcommand{\nb}[2]{
    \fbox{\bfseries\sffamily\scriptsize#1}
    {\sf\small$\blacktriangleright$\textit{#2}$\blacktriangleleft$}
   }
   \newcommand{\cvsversion}{\emph{\scriptsize$-$Id: macro.tex,v 1.9 2005/12/09 22:38:33 giulio Exp $}}
  }
  {\newcommand{\nb}[2]{}
   \newcommand{\cvsversion}{}
  }

\newcommand{\ie}{\textit{i.e.,}\xspace}
\newcommand{\eg}{\textit{e.g.,}\xspace}
\newcommand{\etc}{\textit{etc.}\xspace}
\newcommand{\etal}{\textit{et al.}\xspace}
\newcommand{\etals}{\textit{et al.'s}\xspace}
\newcommand{\aka}{\textit{a.k.a.}\xspace}	
\newcommand{\secref}[1]{Sec.~\ref{#1}\xspace}
\newcommand{\chapref}[1]{Chapter~\ref{#1}\xspace}
\newcommand{\appref}[1]{Appendix~\ref{#1}\xspace}
\newcommand{\figref}[1]{Fig.~\ref{#1}\xspace}
\newcommand{\listref}[1]{Listing~\ref{#1}\xspace}
\newcommand{\equaref}[1]{Equation~\ref{#1}\xspace}
\newcommand{\tabref}[1]{Table~\ref{#1}\xspace}

\newcommand\maddy[1]{{\color{green} \nb{MADDY}{#1}}}
\newcommand\nathan[1]{{\color{blue}\nb{NATHAN}{#1}}}
\newcommand\CARLOS[1]{{\color{orange}\nb{CARLOS}{#1}}}
\newcommand\denys[1]{{\color{purple}\nb{DENYS}{#1}}}
\newcommand\KEVIN[1]{{\color{blue}\nb{KEVIN}{#1}}}
\newcommand\os[1]{{\color{purple}\nb{OSCAR}{#1}}}

\begin{abstract}
Screen recordings are becoming increasingly important as rich software artifacts that inform mobile application development processes. However, the amount of manual effort required to extract information from these graphical artifacts can hinder resource-constrained mobile developers. This paper presents Video2Scenario (\vts), an automated tool that processes video recordings of Android app usages, utilizes neural object detection and image classification techniques to classify the depicted user actions, and translates these actions into a replayable scenario. We conducted a comprehensive evaluation to demonstrate \vts's ability to reproduce recorded scenarios across a range of devices and a diverse set of usage cases and applications. The results indicate that, based on its performance with 175 videos depicting 3,534 GUI-based actions, \vts is able to reproduce $\approx$ 89\% of sequential actions from collected videos. 

\noindent\textbf{Demo {\small URL}:} \texttt{\small \url{https://tinyurl.com/v2s-demo-video}}
\end{abstract}
\section{Introduction}
Rich software artifacts such as crash reports, bug reports, and user reviews provide invaluable information to mobile application developers throughout the development cycle. Recently, it has become increasingly common that screenshots and screen recordings comprise or are used among such artifacts. These graphical artifacts are both easy to obtain and, based on mobile app's heavy reliance on GUI elements to enable functionality, are often better suited to communicate the complex concepts included in a feature request or bug report than a textual description. Unfortunately, while the prevalence of these visual mobile development artifacts continues to rise, the manual effort required to glean relevant information from them remains a challenge. This indicates the need for automated techniques that can extract and analyze data from screen recordings and screenshots.

In this paper, we present \vts, the first Android replay tool that serves to automate the analysis of  video-based mobile development artifacts. \vts functions \textit{solely} on videos; it processes screen recordings of Android app usages, detects and classifies the depicted actions using recently-developed Deep Learning (DL) techniques, and translates these actions into a replayable scenario for a given target device.

We comprehensively analyzed the ability of \vts to reproduce depicted actions and found that it was capable of correctly reproducing $\approx$89\% of the sequential events across 175 collected videos, illustrating its accuracy. We also conducted a case study to evaluate \vts's perceived utility in supporting developers during mobile app development. We found that, across the board, \vts was perceived as a useful tool for app testing and debugging. \vts is an open-source tool that can be found online at \texttt{\small \url{https://tinyurl.com/v2s-tool}}.

\begin{figure*}[t]
    \begin{center}
		\includegraphics[width=0.88\linewidth]{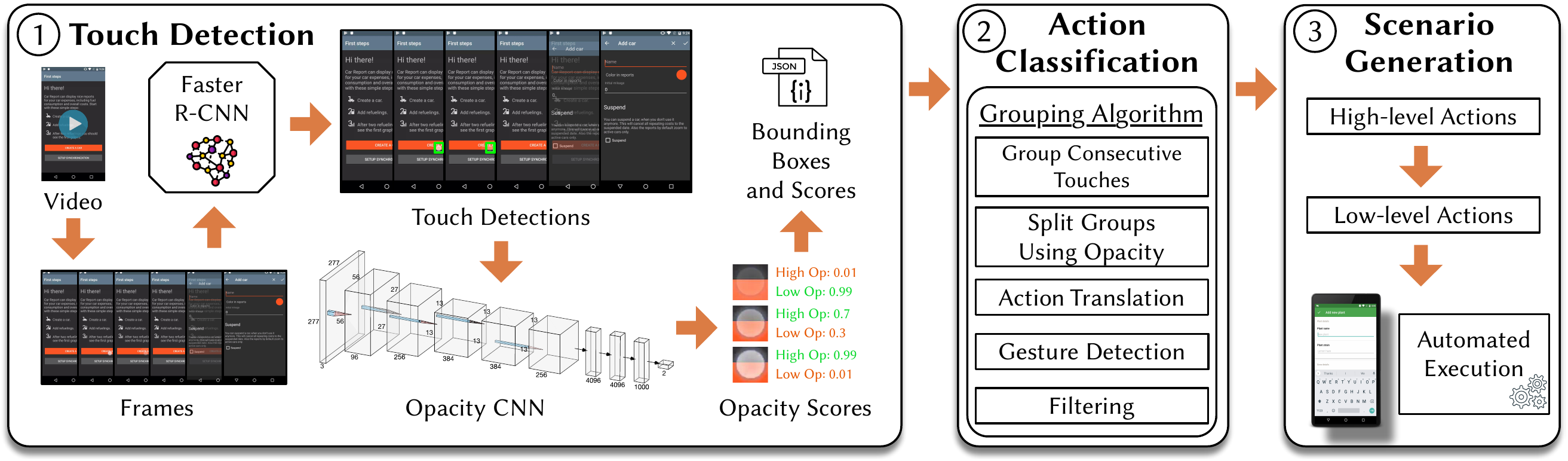}
		\vspace{-0.2cm}
        \caption{\vts Design}
        \label{fig:phases-v2s}
        \vspace{-0.5cm}
    \end{center}
\end{figure*}

\section{\vts Design}
Fig.~\ref{fig:phases-v2s} provides an overview of \vts and its three phases, each of which play a crucial role in accomplishing this functionality. Given that \vts was designed and built with extension in mind, each of these phases and each of the component elements of these phases can be extended or substituted to further the functionality of the pipeline as a whole (\eg users could substitute our object detection model for a custom model). 
The intent of this design choice is to allow researchers and developers to customize \vts for future projects or development use cases.

\vts receives a user-specified configuration file as input that includes the path to the video file to be processed, the path to the object detection and opacity models, and information about the target device. Using the video file's path, \vts enters \hyperref[section:phase1]{Phase 1} of the pipeline, the \textit{video manipulation} and \textit{touch detection} phase, where \vts first extracts each individual video frame and then detects the location and opacity of the touches exhibited in these frames. \vts then enters \hyperref[section:phase2]{Phase 2}, the \textit{action classification} phase, where these detected touches are classified as \texttt{\textbf{\small Tap}}, \texttt{\textbf{\small Long Tap}}, or \texttt{\textbf{\small Gesture}}. Finally, \vts enters \hyperref[section:phase3]{Phase 3}, the \textit{scenario generation} phase, in which a replay script is produced and the input scenario is emulated on the target device. \vts, in its current implementation, can process one video at a time. Details about \vts's algorithms and evaluation can be found in its original research paper \cite{Bernal:ICSE20}.

\subsection{Input Video Specifications} 
\begin{figure}[h]
    \begin{center}
		\includegraphics[width=0.69\columnwidth]{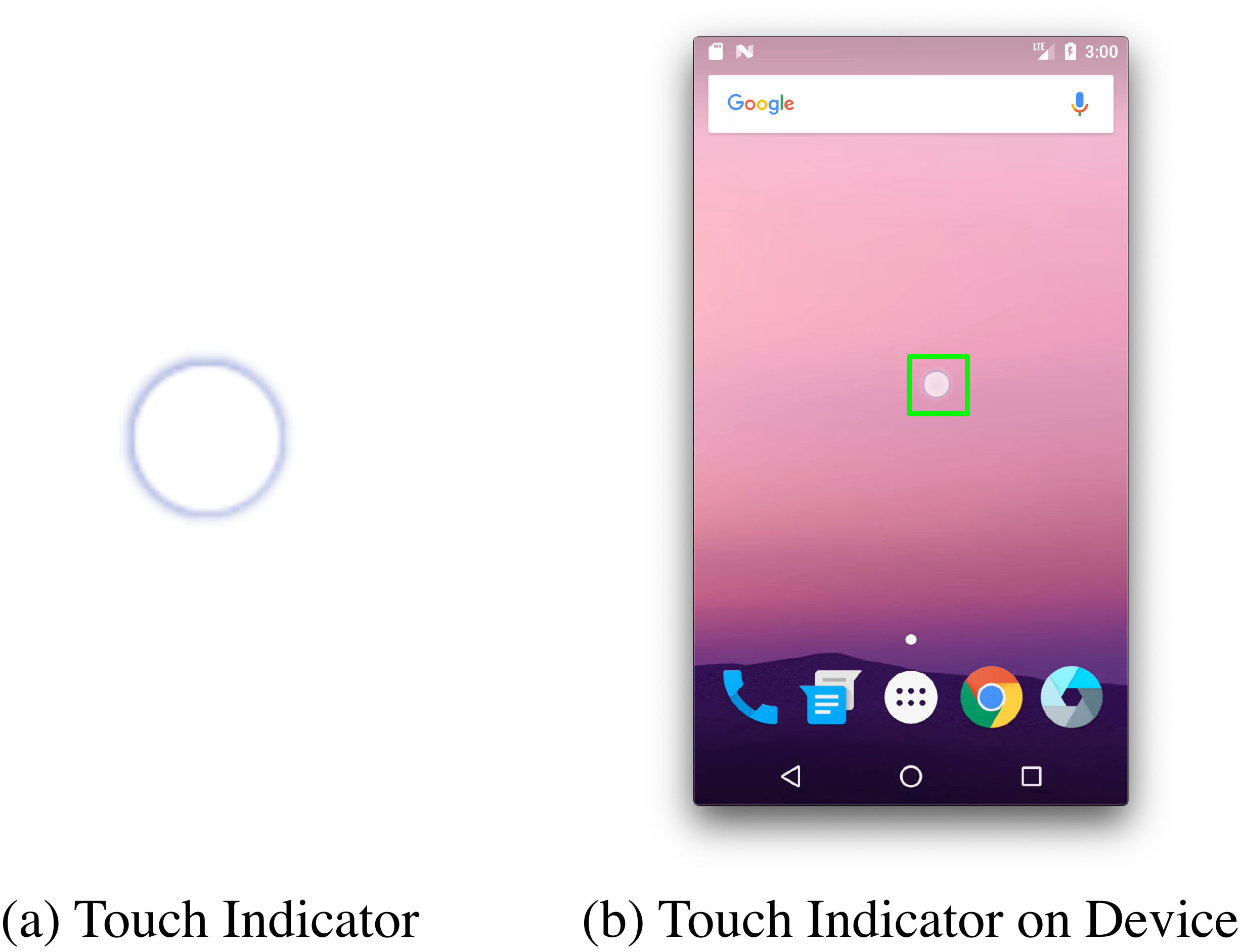}
        \caption{Default Touch Indicator}
        \label{fig:touch}
        \vspace{-0.5cm}
    \end{center}
\end{figure}

Input videos can be captured on a range of devices in one of several different ways, including using the built-in Android \texttt{\textbf{\small screenrecord}} utility or any number of other recording applications available on Google Play~\cite{g-play-recording-apps}. To ensure compatibility with \vts, input videos must adhere to a few specific constraints. Firstly, the frame size of the input video must align with one of our predefined models and equal the screen size of the target Android device. Currently, \vts supports the Nexus 5 and Nexus 6P screen sizes\footnote{Support for additional devices and operating systems can be added, given developers have a connection to the device, the touch indicator image, and device specifications (screen size, etc.). See \cite{Bernal:ICSE20} for more information.}. 
Secondly, \vts requires each input video to be recorded with at least 30 ``frames per second" (FPS) so that fast-paced gestures (such as flicks or rapid swipes) can be accurately detected and replayed. Finally, because \vts aims to detect the location of a user’s finger on the screen, input videos must be recorded with the ``Show Touches" option enabled on the device in developer mode.  This enables an opaque, circular \textit{touch indicator} to appear as the user presses and/or moves their finger on the screen (see Fig.~\ref{fig:touch}.a); as the user lifts their finger to finish an action, the opacity of the indicator decreases. 

\vspace{-.2cm}

\subsection{Phase 1: Video Manipulation and Touch Detection} \label{section:phase1}
Phase 1 first preprocesses the input video and then detects the location of individual touches throughout the video using a Neural Object Detection framework. Phase 1 executes three components: (i) the \textit{FrameExtractor}, (ii) the \textit{TouchDetectorFRCNN}, and (iii) the \textit{OpacityDetector}.  

\subsubsection{FrameExtractor (FE)} To account for different frame rates in input videos, \vts first manipulates each video to adhere to a standard 30 FPS. Then, \vts extracts individual frames from the normalized video using \texttt{\small \textbf{ffmpeg}}~\cite{ffmpeg}.

\subsubsection{TouchDetectorFRCNN (TD)} The \textit{TD} component applies a modified \Faster model (trained utilizing the Tensorflow Object Detection API \cite{TFODA}) to each frame and accurately predicts the location of the bounding box of the touch indicators if any (see Fig.~\ref{fig:touch}.b) \cite{ Ren:PAMI17}.

\subsubsection{OpacityDetector (OD)} After the touches have been localized by the \textit{TD}, the \textit{OD} crops each touch around its bounding box and then feeds these into the \Opacity, which is an extended version of the \AlexNet architecture~\cite{Krizhevsky:NIPS12}. This architecture classifies each touch as high or low opacity. 

Phase 1 then pairs the touch locations detected by the \textit{TD} with the opacity values predicted by the \textit{OD} to form a ``complete" list of detected touch indicators in each video frame. This list is written to a \textit{detection\_full.json} file.

\subsection{Phase 2: Action Classification} \label{section:phase2}
Phase 2 groups the touches detected in Phase 1 and translates them into a set of actions. Phase 2 begins by reading each individual detected touch from the \textit{detection\_full.json} file and, as a preliminary attempt to ensure that each touch that enters the classification process is a true positive, it removes any taps with a confidence measure below 0.7. \vts passes these resulting taps to the \textit{GUIActionClassifier}. 

\subsubsection{GUIActionClassifier (\textit{GAC})} The \textit{GAC} component executes three distinct steps to classify the depicted actions accurately: (i) an action grouping step to organize individual touches across consecutive frames into discrete actions, (ii) a segmentation step to determine the start and end points of complex actions, and (iii) an action translation step which allows for these groupings to be associated with a specific action type. The output of this component is a list of detected actions that can be written to a \texttt{\small \textbf{json}} file.

\noindent \textit{Grouping Consecutive Touches --} As complete actions may occur across consecutive frames, the \textit{GAC} begins by detecting consecutive frames and grouping the detected taps in them into distinct groups based on the timing of the video frames. 

\noindent \textit{Discrete Action Segmentation --} For some fast-paced actions, there may not be separating frames to delineate distinct actions executed by the user. This behavior may be seen when a user is typing on the keyboard or swiping quickly through an article. In these cases, multiple touches will appear in the same frame, and a technique to separate these touches into distinct actions is necessary. The \textit{GAC} component tackles this using a heuristic-based approach, modeling each group of consecutive frames and its associated touches as a connected components graph problem. Each touch within the frame group is considered a node. Nodes are connected if they occur across consecutive frames and are clustered according to pixel-based distance. This means that touches that share similar spatial characteristics across frames are treated as continuous actions.

\noindent \textit{Action Translation --} Actions are translated into one of three predetermined options based on their average location and duration. \texttt{\textbf{\small Taps}} remain in the same relative on-screen location and persist for $<$20 frames. \texttt{\textbf{\small Long Taps}} have similar location requirements but last for $>=$20 frames. Any actions that do not fit into these two categories become \texttt{\textbf{\small Gestures}}. 

Phase 2 outputs a \textit{detected\_actions.json} file containing the list of detected actions with their touches from Phase 1.

\subsection{Phase 3: Scenario Generation} \label{section:phase3}
Phase 3 of \vts harnesses the actions classified in Phase 2 and translates them into a script in the \texttt{\small \textbf{sendevent}} format available in Android's Linux kernel~\cite{getevent}. \vts then utilizes a modified version of the \texttt{RERAN}~\cite{Gomez:ICSE13} binary to replay the predicted scenario on the target device. Phase 3 reads in the actions specified in the \textit{detected\_actions.json} file output by Phase 2 and begins the execution of the \textit{Action2EventConverter}.

\subsubsection{Action2EventConverter (\textit{A2EC})}
The \textit{A2EC} converts the high-level actions produced by Phase 3 into low-level commands in the \texttt{\small \textbf{sendevent}} format. This component utilizes the \texttt{\small \textbf{start\_event}} and \texttt{\small \textbf{end\_event}} commands, and the $x$ and $y$ coordinates of the various actions to generate the replay script and control the UI behavior on the target device. 

Each action begins with the \texttt{\small \textbf{start\_event}} command to indicate the start of an event. For the \texttt{\textbf{\small Tap}} action, the \textit{A2EC} component calculates the centroid or average location. For the \texttt{\textbf{\small Long Tap}} action, the \textit{A2EC} component utilizes this same centroid point but sustains this command for a specified duration of time, depending on the number of frames for which the action persists. Finally, for \texttt{\textbf{\small Gesture}} actions, the \textit{A2EC} component iterates over and appends each $(x,y)$ coordinate pair of the action to the script. For each of these actions, the \textit{A2EC} terminates the action by appending the \texttt{\small \textbf{end\_event}} command to the script.

The \textit{A2EC} allots a duration of 33 milliseconds per frame and calculates an appropriate timestamp for each command appended to the script. As 33 milliseconds is the delay between frames for a video at 30 FPS, this ensures proper timing of actions to be executed in the generated script.

\subsubsection{Scenario Replay} 
The script generated by the \textit{A2EC} is written to a \textit{send\_events.log} file. This log file is then converted by the \textit{Translator} component into a format that is executable on the target device. Once this translation has occurred, \vts pushes the executable file and a modified version of the \texttt{RERAN} binary to the device, starts a screen recording of the generated scenario, and executes the V2S script. 

\section{\vts Evaluation}
In order to evaluate the capabilities of \vts as a whole, we asked five different research questions and conducted associated studies. We measured: (i) the accuracy of the touch detection \Faster model; (ii) the accuracy of the opacity detection \AlexNet model; (iii) the accuracy of \vts on different usage scenarios; (iv) the runtime performance of \vts; and (v) the practical utility of \vts. The results here are summarized from the original \vts paper~\cite{Bernal:ICSE20}.

\subsection{Accuracy of \Faster}
\subsubsection{Design} 
In order to evaluate the \Faster model’s ability to correctly identify the location of a touch indicator within an image, we generated a dataset of 15,000 images containing touch indicators and split this data 70\%-30\% into training and testing sets, respectively. For this preliminary study, we trained two different models for two different screen sizes (the Nexus 5 and the Nexus 6P) using the Tensorflow Object Detection API~\cite{TFODA}. More information about this training process can be found in the \vts paper \cite{Bernal:ICSE20}.

For this study, we considered the model’s \textit{Mean Average Precision} (mAP), which is a commonly-used metric to determine accuracy for object detection tasks \cite{Bernal:ICSE20}.
This metric is calculated as $mAP = TP/(TP+FP)$, where a $TP$ is a region correctly identified as a touch indicator and an $FP$ is a region that is incorrectly marked as containing a touch indicator. We also considered the recall metric of this model to determine the frequency with which our model misses a touch indicator when one is present.

\subsubsection{Results}
All of the models achieved $\approx$ 97\% mAP in predicting the location of the touch indicator. The model was also able to achieve $\approx$ 99\% recall, meaning that the model rarely missed the detection of the touch indicator.

\subsection{Accuracy of \Opacity}
\subsubsection{Design}
We evaluated the \Opacity model's accuracy in correctly classifying whether a touch indicator has a high or low opacity. We generated a dataset of 10,000 images containing equal numbers of high-opacity and low-opacity touch indicators and then split these images 70\%-30\% into training and testing datasets, respectively. We used TensorFlow and Keras to create and train a modified \Opacity. 

For this study, we considered the \textit{Mean Average Precision} (mAP) of this model, where a $TP$ is an indicator that was identified as having the correct opacity, and an $FP$ is an indicator that was predicted to have the incorrect opacity value.

\subsubsection{Results}
We found that the precision of this model was $\approx$ 98-99\% mAP. This indicates that the opacity model is very accurate in its ability to correctly distinguish between high and low-opacity touch indicators. 

\subsection{Accuracy on Usage Scenarios}
\subsubsection{Design}
We designed two studies to test \vts's ability to replicate an original usage scenario depicted in a video recording. The \textit{Controlled Study} (\textit{CS}) was meant to ensure the \textit{depth} of \vts's abilities in reproducing a variety of bug crashes, synthetically-injected and real application failures, and normal usage scenarios. Then, in the \textit{Popular Applications Study} (\textit{PAS}), we assessed the \textit{breadth} of \vts in its ability to accurately reproduce a variety of usage scenarios depicted in a diverse set of applications from the Google Play store.

In the \textit{CS}, eight participants were recruited from William \& Mary 
to record eight different scenarios each. These scenarios came from two of the following four categories: (i) normal usage cases, (ii) bugs in open source apps, (iii) real crashes, and (iv) synthetically-injected crashes in open source apps. Four participants recorded on the Nexus 5 and four recorded on the Nexus 6P. The participants familiarized themselves with the scenario before recording. Each of the buggy scenarios was extracted from previous studies of a similar nature~\cite{Chaparro:FSE'19, Moran:FSE15, Moran:ICST16}. 

For the \textit{PAS}, we downloaded the top-two rated applications from 32 categories in the Google Play store. Two of the authors then recorded two scenarios per application, with each scenario differently exhibiting one of the major features of the application. To ensure that \vts could reproduce videos on different devices, one author recorded their scenarios on the Nexus 5 and the other on the Nexus 6P.

To verify \vts' accuracy, we manually determined the ground truth action sequences and used this to compute four different metrics: (i) Levenshtein distance, (ii) Longest Common Subsequence (LCS), (iii) precision and recall, and (iv) manual video comparison. Levenshtein distance is a metric that depicts the number of alterations necessary to transform one sequence into another. The LCS metric represents the longest sequence of continuous matching actions produced by \vts when compared to the ground truth sequence. To compute the precision and recall, we created an unordered ``action pool" for each scenario in our studies and for each predicted action type. Comparing this predicted ``action pool" to the ground truth ``action pool" allowed us to calculate these metrics overall and per individual action type. Finally, each scenario was manually reviewed and marked as successful as long as the reproduced behavior exercised the  same overall functionality as the original. 

\subsubsection{Results}
We briefly summarize the results.
\paragraph{Levenshtein Distance} For the \textit{CS} and \textit{PAS}, the avg. distance value was 0.85 and  1.17 changes, respectively.
\paragraph{LCS} In the \textit{CS}, \vts was able to match 95.1\% of the sequential actions, and for the \textit{PAS}, \vts was able to correctly emulate 90.2\% of these consecutive events. 
\paragraph{Precision and Recall}  
Overall, \vts had very high precision and recall values for all event types: 
for the \textit{CS}, \vts achieved 95.3\% precision and 99.3\% recall, and for the \textit{PAS}, \vts achieved 95\% precision and 97.8\% recall. 
\paragraph{Manual Video Comparison} \vts accurately replicated 93.75\% of the 64 scenarios in the  \textit{CS} and 94.48\% of sequential actions. For the \textit{PAS}, of the 111 videos fed into \vts, the tool accurately replayed 81.98\% of scenarios and 89\% of sequential actions. 

\subsection{\vts Runtime Performance} 
\subsubsection{Design} 
To assess the execution performance of \vts, we calculated the average runtime of the pipeline per video. 

\subsubsection{Results} 
We measured the performance per frame (in seconds per frame or $s/f$) of each major step that makes up \vts. We determined the average runtime of (i) the frame extraction process (0.045 $s/f$); (ii) the touch detection process (1.09 $s/f$); and (iii) the opacity classification step (0.032 $s/f$). By our calculations, an average 3-minute video processed by \vts in full would take $\approx$ 105 minutes. This performance could be easily enhanced by parallelizing the computation.

\subsection{Industrial Utility} 
\subsubsection{Design}
We wanted to understand if mobile developers perceive \vts as useful. We interviewed three software engineers to assess the potential role of \vts in their day-to-day operations. The first section of the interview aimed at gaining an understanding of the developer’s backgrounds and the tools that they use to accomplish their every-day tasks. The second section of the interview was intended to allow participants to assess the performance and utility of \vts. This was accomplished by presenting them with a demonstration of the input and replication videos and the resulting detections, action list, and replay script that \vts produces.

\subsubsection{Results}
All of the participants agreed that \vts is highly accurate at reproducing actions depicted in an input video. Each of the participants also viewed \vts as a potentially useful tool for app testing and debugging.

\section{Final Remarks \& Future Work}

In this tool demonstration paper, we have presented \vts, a novel tool for translating video recordings of mobile app usages into replayable scenarios. \vts facilitates different testing and debugging activities by allowing for easy replay of app usages and bugs captured via screen recordings, and our evaluation illustrates its effectiveness across several dimensions. As future work, we plan to add support for multi-fingered actions and train additional object detection models so that \vts can be used with a more diverse set of devices. 

\section{Acknowledgements.} 

This research was supported in part by the NSF CCF-1955853 and CCF-1815186 grants. 
Any opinions, findings, and conclusions expressed herein are the authors’ and do not necessarily reflect those of the sponsors.

\bibliographystyle{ieeetr}
\bibliography{references}
\end{document}